\documentclass[prodmode,permissions]{acmsmall-ec16}
\usepackage[dvipsnames,usenames,table]{xcolor}
\usepackage[numbers,sort&compress]{natbib} 

\usepackage{graphicx}
\usepackage{transparent}
\usepackage{amsmath,amsfonts,amssymb}
\usepackage[noend]{algpseudocode}
\usepackage{algorithm}
\usepackage{tikz}

\usepackage[colorlinks=true,urlcolor=blue,linkcolor=RoyalBlue,citecolor=OliveGreen]{hyperref}

\newcommand{\suppress}[1]{}

\newcommand{\pp}{\mbox{\boldmath $p$}}

\newcommand{\dt}{\mbox{$\delta$}}

\newcommand{\seg}{\mbox{\rm seg}}
\newcommand{\segs}{\mbox{\rm segments}}
\newcommand{\valuex}{\mbox{\rm value}}
\newcommand{\rate}{\mbox{\rm rate}}

\newcommand{\good}{\mbox{\rm good}}

\newcommand{\ra}{\rightarrow}

\newcommand{\Q}{\mbox{\rm\bf Q}}
\newcommand{\Qplus}{\Q^+}

\newcommand{\CM}{\mbox{${\cal M}$}}

\newcommand{\CCP}{\mbox{${\cal CP}$}}

\newcommand{\R}{\mbox{\rm\bf R}}
\newcommand{\Rplus}{\R_+}

\makeatletter
\def\fnum@figure{{\bf Figure \thefigure}}
\def\fnum@table{{\bf Table \thetable}}

\long\def\@mycaption#1[#2]#3{\addcontentsline{\csname
 ext@#1\endcsname}{#1}{\protect\numberline{\csname
  the#1\endcsname}{\ignorespaces #2}}\par
     \begingroup
       \@parboxrestore
          \small
       \@makecaption{\csname fnum@#1\endcsname}{\ignorespaces
#3\endgroup}
      }

\newcommand{\SB}{c}

\newcommand{\UB}{d}

\begin{document}

\title{New Convex Programs for Fisher's Market Model \\
and its Generalizations\footnote{Supported by NSF Grant CCF-1216019}}

\author{NIKHIL R. DEVANUR \footnote{Microsoft Research. nikdev@microsoft.com}\\
KAMAL JAIN\footnote{Faira Inc. kamaljain@gmail.com}\\
TUNG MAI\footnote{College of Computing,
Georgia Institute of Technology. maitung89@gatech.edu}\\
VIJAY V. VAZIRANI\footnote{College of Computing,
Georgia Institute of Technology. vazirani@cc.gatech.edu}\\
SADRA YAZDANBOD\footnote{College of Computing,
Georgia Institute of Technology. syazdanb@cc.gatech.edu}
}
%

\begin{abstract}
We present the following results pertaining to Fisher's market model:
\begin{itemize}

\item
We give two natural generalizations of Fisher's market model: In model $\CM_1$, sellers can declare an upper bound
on the money they wish to earn (and take back their unsold good), and in model $\CM_2$, buyers can declare an upper bound
on the amount to utility they wish to derive (and take back the unused part of their money). 
\item
We derive convex programs for the linear case of these two models by generalizing a convex program due to Shmyrev and the
Eisenberg-Gale program, respectively.
\item
We generalize the Arrow-Hurwicz theorem to the linear case of these two models, hence deriving alternate convex programs.
\item
For the special class of convex programs having convex objective functions and {\em linear} constraints, we
derive a simple set of rules for constructing the dual program (as simple as obtaining the dual of an LP).
Using these rules we show a formal relationship between the two seemingly different convex programs for linear Fisher markets, due to
Eisenberg-Gale and Shmyrev; the duals of these are the same, upto a change of variables.
\end{itemize}
\end{abstract}

\date{}
\maketitle

\section{Introduction}
\label{sec.intro}

The fundamental market model of Fisher has been widely studied in economics \cite{scarf,gale} and in theoretical computer
science \cite{ch6AGT,DPSV,JVEG,CV,CMV,KJMM,GMSV,GMV,ch5AGT}. A remarkable convex program, due to 
\citet{eisenberg}, captures equilibrium allocations, and its dual variables capture equilibrium prices for this model for several utility functions including linear, CES and Leontief utilities \cite{gale,CV,ch6AGT}. 

Another seemingly very different convex program, $\CCP$, for the same market was discovered by \citet{shmyrev2009algorithm}: 
whereas variables in the 
Eisenberg-Gale convex program represent allocations of goods received by each buyer, variables in $\CCP$ are prices of goods and the amount of money spent by each agent on each good. 
We show how these two convex programs are related:
one can define a dual convex program for each of these, 
and these two duals are the same, upto a change of variables.

We next present two natural generalizations of 
Fisher's market model: In model $\CM_1$, sellers can declare an upper bound
on the money they wish to earn (and take back their unsold good), and in model $\CM_2$, buyers can declare an upper bound
on the amount to utility they wish to derive (and take back the unused part of their money); see Section \ref{sec.model}
for formal definitions.

Model $\CM_2$ is clearly natural: it is reasonable to assume that a buyer would only want to buy goods that are
necessary, i.e., give a certain bounded amount of utility, and she desires to keep the rest of her money for future use.
Our model $\CM_1$ is also natural, in particular in the following situation. Assume that each seller is selling his services in the
market. 
In the last half century, society has seen the emergence of a multitude of very high end jobs which call for a lot of expertise and in turn pay very large salaries.
Indeed, the holders of such jobs do not need to work full time to make a comfortable living and one sees  numerous such people preferring to work for
shorter hours and having a lot more time for leisure. High end dentists, doctors and investors fall in this category. 
Our model $\CM_1$ allows such agents to specify a limit on their earnings beyond which they do not wish
to sell their services anymore. 

Observe that if many agents desire $j$'s services and 
$j$ has a low upper bound on his earnings, say $\SB_j$, then the price, $p_j$ of this good will be set high enough so that the 
buyers demand very little of his services. In particular, he will sell only $\SB_j/p_j$ fraction of his services. Thus, in model $\CM_1$, the supply of a good is a function of the prices. 
Similarly, in $\CM_2$, the amount of money a buyer spends in the market is a function of the prices.
In the presence of these additional constraints, do equilibria exist and can they be computed in polynomial time? 

In this paper, we go further: we give convex programs that capture equilibria for each of the models.
Generalizations of $\CCP$ and the 
Eisenberg-Gale programs, respectively, yield convex programs for $\CM_1$ and $\CM_2$ for the case of linear utilities.
Existence of equilibria for both models follows from these convex programs.
We further show that both models admit rational equilibria, i.e., prices and allocations are rational numbers if all parameters specified in the instance are rational. As a consequence, the ellipsoid algorithm will find a solution to the convex programs in
polynomial time. In addition, for $\CM_2$, we also give convex programs for Leontief and CES (for parameter values 
$0 < \rho < 1$) utility functions. 
Finally, we show that for market $\CM_1$ under spending constraint utilities, 
a generalization of $\CCP$ yields a convex program.

A completely different way of deriving convex programs
for markets follows from the classic Arrow-Hurwicz theorem \cite{ArrowH}, which applies to Arrow-Debreu markets with weak
gross substitutes (WGS) utilities (this includes linear utilities). We present extensions of the Arrow-Hurwicz's theorem to markets
$\CM_1$ and $\CM_2$ for the linear case and hence derive alternative convex programs for these markets.\footnote{The solution of these convex programs requires additional properties, 
such as finding an initial bounding box. These details are in Appendix \ref{sec:ellipsoid}.} We note that
the Arrow-Hurwicz theorem applies to utility functions satisfying weak gross substitutability (WGS), which includes linear utilities.
Our result opens up the possibility that one could extend this approach to all WGS utilities for $\CM_1$ and $\CM_2$. 

Our paper also contributes to convex programming duality theory\footnote{This part of the current paper is incorporated
from the following unpublished manuscript:\\
N. R. Devanur. Fisher Markets and Convex Programs. Manuscript, 2010.}.
This duality is usually stated in its most general form, with convex objective functions and convex constraints, e.g., see the
excellent references by  \citet{BV} and \citet{rockafellar1970convex}. However, 
at this level of generality the process of constructing the dual of a convex program is a little tedious,
in contrast to LP duality where there is a simple set of rules for obtaining the dual of an LP.
In this paper, we consider a special class of convex programs, those with convex objective functions and {\em linear} constraints,
and derive a simple set of rules to construct the dual,\footnote{The dual is obtained using the usual Lagrangian relaxation technique. We show a ``short-cut" for applying this technique, making it especially easy to derive the dual for the special case we consider. } 
which is almost as simple as the LP case.

 These rules have found further applications:  to derive convex programs for Fisher markets under spending constraint utilities\footnote{See Section \ref{sec.model} for a definition} \cite{BDXnikhil}, for Fisher markets with transaction costs \cite{CDK10}, and for Arrow-Debreu market with linear utilities \cite{devanur2013rational}.
 They have been used in the design of algorithms: 
 for simplex like algorithms for spending constraint utilities and perfect price discrimination markets \cite{garg2013towards},
 in analyzing the convergence of the Tatonnement process \cite{cheung2013tatonnement}, 
 in designing online algorithms for scheduling \cite{im2014selfishmigrate,buchbinder2014online,devanur2014primal}, 
 and online algorithms for welfare maximization with production costs \cite{huang2015welfare}. They have also been used in bounding the price of anarchy of certain games \cite{kulkarni2015robust}.

\section{Models}
\label{sec.model}

Fisher's model is the following: let $A$ be a set of $n$ divisible goods and $B$ be a set of $m$ buyers. 
Each buyer $i$ comes to the market with money $m_i$, and we may assume
w.l.o.g. that the market has one unit of each good. Each buyer $i$ has a utility function,
$f_i: \Rplus^m \ra \Rplus$, giving the utility that $i$ derives from each bundle of goods.
Prices $p$ and allocations $x$ are said to form an {\em equilibrium} if each buyer $i$ gets an optimal bundle
of goods, and each good having non-zero price is fully sold. Clearly, in an equilibrium, each agent will fully
spend her money. For convenience, we will assume that each good is sold by a unique seller.

Utility function for buyer $i$ $f_i: \Rplus^m \ra \Rplus$ is said to be {\em linear} if
there are parameters $u_{ij} \in \Rplus$, specifying the utility derived by $i$ from one unit of good $j$.
Her utility for the entire bundle is additive, i.e., $f_i(x) = \sum_{j \in B} {u_{ij} x_{ij}}$.
Utility function $f_i$ is said to be {\em Leontief} if,
given parameters $a_{ij} \in \Rplus \cup \{0\}$ for each good $j \in B$,  
 $f_i(x) = \min_{j \in B} {x_{ij}/a_{ij}}$.
Finally, $f_i$ is said to be {\em constant elasticity of substitution (CES) with parameter $\rho$} if 
given parameters $\alpha_j$ for each good $j \in B$,  
\[f_i(x) = \left( \sum_{j=1}^n \alpha_j x_j^{\rho} \right)^{{1 \over {\rho}}}  . \]

Our two generalizations of Fisher's model are the following. In the first model, $\CM_1$, each seller $j$ has an upper bound 
$\SB_j$ on the amount of money $j$ wants to earn in
the market. Once he earns $\SB_j$, selling the least amount of his good, he wants to take back the unsold
portion of his good. In equilibrium, buyers spend all their money and get an optimal bundle of goods.
In the second model, $\CM_2$, buyers 
have upper bounds $\UB_i$ on the utility they want to derive in the market. Once buyer $i$ derives utility $\UB_i$,
spending the least amount of money at prices $p$, she wants to keep the left-over money. Once again,
in equilibrium, each good with a positive price should be fully sold. 

We next define the spending constraint model.
As before, let $A$ be a set of divisible goods and $B$ a set of buyers, $|A| = n, \ |B| = m$.
Assume that the goods are numbered from 1 to $n$ and the buyers are numbered
from 1 to $m$. 
Each buyer $i \in B$ comes to the market with a specified amount of
money, say $m_i \in \Qplus$, and we are specified the quantity, $b_j \in \Qplus$ of
each good $j \in A$. 
For $i \in B$ and $j \in A$, let $f_j^i: \ [0, m_i] \rightarrow \Rplus$
be the {\em rate function} of buyer $i$ for good $j$; it specifies the rate
at which $i$ derives utility per unit of $j$ received, as a function of
the amount of her budget spent on $j$. If the price of $j$ is fixed at
$p_j$ per unit amount of $j$, then the function $f_j^i/p_j$
gives the rate at which $i$ derives utility per dollar spent, as a function of
the amount of her budget spent on $j$. Define $g_j^i: [0, m_i] \ra \Rplus$ as follows:
\[ g_j^i(x) = \int_0^x { {{f_j^i(y)} \over p_j} dy }  .\]
This function gives the utility derived by $i$ on spending $x$ dollars on good $j$ at price $p_j$.

In this paper, we will deal with the case that $f_j^i$'s
are decreasing step functions. If so, $g_j^i$ will be a piecewise-linear and concave
function. The linear version of Fisher's problem \cite{scarf} is the special
case in which each $f_j^i$ is the constant function so that $g_j^i$ is a linear function.
Given prices $\pp = (p_1, \ldots, p_n)$ of all goods, each buyer wants a utility maximizing bundle of goods.
Prices $\pp$ are equilibrium prices if each good with a positive price is fully sold.

\newcommand{\tom}[1]{{\bf #1}}
\newcommand{\fixme}{\text{\rm FIX ME}}
\newcommand{\Exp}{{\mathbf{E}}}
\newcommand{\Prob}{{\mathbf{Pr}}}
\newcommand{\BBB}{{\mathcal{B}}}

\newcommand{\lal}[1]{}
\newcommand{\aij}{ a_{ij}}
\newcommand{\xij}{ x_{ij}}
\newcommand{\uij}{ u_{ij}}
\newcommand{\bij}{ b_{ij}}
\newcommand{\cij}{ c_{ij}}
\newcommand{\etaij}{{\eta_{ij}}}
\newcommand{\Dt}{{\tilde{D}}}
\newcommand{\Pt}{{\tilde{P}}}
\newcommand{\Xt}{{\tilde{X}}}
\newcommand{\xtij}{{\widetilde{x}_{ij}}}
\newcommand{\utij}{{\widetilde{u}_{ij}}}
\newcommand{\alts}{{\tilde{\alpha}^*}}
\newcommand{\alphahat}{\widehat{\alpha}}
\newcommand{\G}{\Omega}
\newcommand{\bmax}{{b_{\max}}}

\newcommand{\thetaij}{\theta_{ij}}
\newcommand{\lt}{\tilde{L}}
\newcommand{\pt}{\tilde{P}}
\newcommand{\ptj}{\tilde{p}_j}
\newcommand{\rti}{\tilde{R}_i}
\newcommand{\als}{\alpha^{*(l)}}
\newcommand{\algl}{\alg^{(l)}}
\newcommand{\tl}{T^{(l)}}
\renewcommand{\sl}{S^{(l)}}
\newcommand{\error}{\epsilon \opt}
\newcommand{\errort}{{~\textsf{error 2}~}}
\newcommand{\optrate}{~\textsf{optimum rate}~}
\newcommand{\OB}[1]{O\left(#1\right)}
\newcommand{\rb}[1]{R_{(#1)} }
\newcommand{\tb}[1]{t_{(#1)} }
\newcommand{\Ib}[1]{I_{(#1)} }
\newcommand{\Bb}[1]{B_{(#1)} }
\newcommand{\Pb}[1]{P_{(#1)} }
\newcommand{\dtb}[1]{\dt_{(#1)} }
\newcommand{\del}[1]{\Delta_{(#1)} }
\newcommand{\grid}{\Lambda}
\newcommand{\blnnd}{\mathcal{B}}
\newcommand{\sqkm}{\sqrt{\frac{k}{m}}}
\newcommand{\sqmk}{\sqrt{\frac{m}{k}}}

\section{Convex programming duality}

\subsection{Conjugate}
We now define the {\em conjugate} of a function,  and note some of the properties; 
see \citet{rockafellar1970convex} for a detailed treatment.
This will be the key ingredient to extend the simple set of rules for LP duality to convex programs.
\newcommand{\grad}{\nabla}
\newcommand{\muij}{\mu_{ij}}
Suppose that $f: \mathbb{R}^n \rightarrow \mathbb{R}$ is a function.
The conjugate of $f$ is  $f^*:\mathbb{R}^n \rightarrow\mathbb{R} $ and is defined as follows:
 \[ f^*(\mu) := \sup_x\{ \mu^Tx - f(x) \}. \]
Although the conjugate is defined for any function $f$, for the rest of the article, we will assume
that $f$ is {\em strictly convex and differentiable}, since this is the case that is most interesting to the applications we discuss.

\noindent {\bf Properties of $f^*$:}
\begin{itemize}
  \item[$\bullet$] $f^* $ is strictly convex and differentiable. (This property holds even if $f$ is not strictly convex and differentiable.)
  \item[$\bullet$] $f^{**} = f$. (Here we use the assumption that $f$ is strictly convex and differentiable.)
  \item[$\bullet$] If $f$ is separable, that is $f(x) = \sum_i f_i(x_i)$, then $f^*(\mu) =\sum_i f_i^*(\mu_i)$.
  \item[$\bullet$] If $g(x) = cf(x)$ for some constant $c$, then $g^*(\mu) = c f^*(\mu/c)$.
  \item[$\bullet$] If $g(x) = f(cx)$ for some constant $c$, then $g^*(\mu) = f^*(\mu/c)$.
  \item[$\bullet$] If $g(x) = f(x+ a)$ for some constant $a$, then $g^*(\mu) = f^*(\mu) - \mu^T a$.
  \item[$\bullet$] If $\mu$ and $x$ are such that $f(x) + f^*(\mu) = \mu^Tx$ then
$\grad f(x) = \mu $ and $\grad f^*(\mu) = x$.
\item[$\bullet$] Vice versa, if $\grad f(x) = \mu $ then $\grad f^*(\mu) = x$ and
$f(x) + f^*(\mu) = \mu^Tx$.
\end{itemize}
We say that $(x,\mu)$ form a complementary pair wrt $f$ if they satisfy one of the last two conditions stated above.
We now calculate the conjugates of some simple strictly convex and differentiable functions. These will be useful later.
\begin{itemize}
\item[$\bullet$] If $f(x) = \frac{1}{2} x^2$, then $\nabla f (x) = x$. Thus $f^*(\mu) $ is obtained by letting
$\mu = x$ in $\mu^Tx - f(x)$, which is then equal to $ \frac{1}{2} \mu^2$.
\item[$\bullet$] If $f(x) = -\log(x)$, then $\nabla f (x) = -1/x$. Set $\mu = -1/x$ to get $f^*(\mu) = -1 + \log(x) = -1 - \log(-\mu)$.
\item[$\bullet$] Suppose $f(x) = x \log x$. Then
$\nabla f(x) = \log x + 1 = \mu$. So $x = e^{\mu -1}$.
$f^*(\mu) = \mu x - f(x) = x (\log x +1 ) - x \log x = x = e^{\mu-1}$.
That is, $f^*(\mu)  = e^{\mu-1}$.

\end{itemize}

\subsection{Convex programs with linear constraints}
Consider the following (primal) optimization problem.
\[\max \textstyle \sum_i c_i x_i - f(x) \text{ s.t. }\]
\[ \forall~ j, \textstyle \sum_i \aij x_i \leq b_j. \]
We will derive a minimization problem that is the {\em dual} of this, using Lagrangian duality.
This is usually a long calculation. The goal of this exercise is to identify a shortcut for the same.
Define the Lagrangian function
\[L(x,\lambda) :=  \sum_i c_i x_i - f(x) +\sum_j \lambda_j
(b_j -  \sum_i \aij x_i ).\]
We say that $x$ is feasible if it satisfies
all the constraints of the primal problem.
Note that for all $\lambda \geq 0 $ and $x$ feasible,
$L(x,\lambda) \geq \sum_i c_i x_i - f(x) $.
Define the dual function
\[ g(\lambda) = \max_x L(x,\lambda).\]
So for all $\lambda,x$,  $g(\lambda) \geq L(x,\lambda).$
Thus
$\min_{\lambda\geq 0} g(\lambda) $ is an upper bound on the optimum value for the primal program.
The dual program is essentially $\min_{\lambda\geq 0} g(\lambda)$. We further simplify it as follows.
Letting  $\mu_i = c_i - \sum_j \aij \lambda_j,$ we can rewrite the expression for $L$ as
\[L = \sum_i \mu_i x_i - f(x) + \sum_j b_j \lambda_j.\]
Now note that $g(\lambda) = \max_x L(x,\lambda) = \max_x \{\sum_i \mu_i x_i - f(x)\} + \sum_j b_j \lambda_j = f^*(\mu) + \sum_j b_j \lambda_j$.
Thus we get the dual optimization problem:
\[\min \textstyle  \sum_j b_j\lambda_j + f^*(\mu) \text{ s.t.} \]
\[\forall~ i, \textstyle \sum_j \aij \lambda_j  = c_i - \mu_i,\]
\[\forall~ j, \lambda_j \geq 0 .\]
Note the similarity to LP duality. The differences are as follows.
Suppose the concave part of the primal objective is  $-f(x)$. There is an extra variable $\mu_i$ for every
variable $x_i$ that occurs in $f$. In the constraint corresponding to $x_i$, $-\mu_i$ appears on the RHS along with the constant term.
Finally the dual objective has $f^*(\mu)$ in addition to the linear terms. In other words, we {\em relax}  the constraint
corresponding to $x_i$ by allowing a slack of $\mu_i$, and {\em charge} $f^*(\mu)$ to the objective function.

Similarly, the primal program with non-negativity constraints on variables has the following dual program. 

\begin{minipage}{0.35\textwidth}  
	\[\text{Primal: }\max \textstyle \sum_i c_i x_i - f(x) \text{ s.t. }\]
\[ \forall~ j,\textstyle  \sum_i \aij x_i \leq b_j,\]
\[ \forall~ i, x_i \geq 0 .\]
\end{minipage}
\hfill\vline\hfill
\begin{minipage}{0.35\textwidth}  
\[\text{Dual: } \min \textstyle \sum_j b_j\lambda_j + f^*(\mu) \text{ s.t.} \]
\[\forall~ i,\textstyle  \sum_j \aij \lambda_j \geq c_i - \mu_i,\]
\[\forall~ j, \lambda_j \geq 0 .\]
\end{minipage}

As we saw, the optimum for the primal program is lower than the optimum
for the dual program (weak duality).
In fact, if the primal constraints are strictly feasible, that is there exist $x_i$ such that for all $j$  $\sum_i \aij \xij < b_j$,
then the two optima are the same (strong duality) and the following  generalized complementary slackness conditions characterize them:
\begin{itemize}
\item $x_i > 0 \Rightarrow \sum_j\aij \lambda_j= c_i - \mu_i$,
\item $\lambda_j > 0 \Rightarrow \sum_i \aij x_i = m_i$ and
\item $x$ and $\mu$ form a complementary pair wrt $f$, that is,
$\mu = \grad f(x), x = \grad f^*(\mu)$ and $f(x) + f^*(\mu) = \mu^Tx$.
\end{itemize}
Similarly the dual of a minimization program has the following form. 

\begin{minipage}{0.35\textwidth}  
	\[\textstyle \text{Primal: }\min \sum_i c_i x_i + f(x) \text{ s.t. }\]
\[ \textstyle \forall~ j, \sum_i \aij x_i \geq b_j,\]
\[ \forall~ i, x_i \geq 0 .\]
\end{minipage}
\hfill\vline\hfill
\begin{minipage}{0.35\textwidth}  
\[\textstyle \text{Dual: }\max \sum_j b_j\lambda_j - f^*(\mu) \text{ s.t.} \]
\[\textstyle \forall~ i, \sum_j \aij \lambda_j \leq c_i + \mu_i,\]
\[\forall~ j, \lambda_j \geq 0 .\]
\end{minipage}

\subsection{Infeasibility and Unboundedness}
When an LP is infeasible,  the dual becomes unbounded. The same happens with these convex programs as well. 
We now give the proof for some special cases. 
Suppose first that the set of linear constraints is itself infeasible, that is, there is no solution to the set of inequalities 
\begin{equation}\label{eq.feasible} \forall~ j, \sum_i \aij x_i \leq b_j.\end{equation}
Then by Farkas' lemma, we know that there exists numbers $\lambda_j\geq 0$ for all $j$ such that 
\[ \forall~ i, \sum_j \aij \lambda _j = 0, \text{and} \sum_j \lambda_j b_j < 0.\]
Now $g(\lambda) = f^*(c) +  \sum_j \lambda_j b_j $, and by multiplying all the $\lambda_j$ 
by a large positive number, $g$ can be made arbitrarily small. 

Now suppose that the feasible region defined by the inequalities (\ref{eq.feasible}) and the domain of $f$ defined 
as $ dom(f) = \{x: f(x) < \infty \}$ are disjoint. Further assume for now that $f^*(c) < \infty$ and that 
there is a strict separation between the two, meaning that for all $x $ feasible and $y \in dom(f)$,  
$d(x,y) > \epsilon$ for some $\epsilon > 0$. 
Then once again by Farkas' lemma we have that there exist $\lambda_j\geq 0$ for all $j$ and $\delta > 0$ 
such that 
\[\forall y \in dom(f),   \sum_{i,j} \aij \lambda _j y_i  >  \sum_j \lambda_j b_j (1 + \delta) .\]
This implies that $g(\lambda) < f^*(c) -  \delta \sum_j \lambda_j b_j $, and as before, 
by multiplying all the $\lambda_j$ by a large positive number,  $g$ can be made arbitrarily small.

\section{Convex programs for Fisher markets}
\label{sec.4}

%
The following is the classic Eisenberg-Gale convex program for Fisher markets with linear utilities.
An optimum solution to this program captures equilibrium allocation for the corresponding market.
\[ \textstyle\max \sum_{i} m_i \log u_i \text{ s.t.} \]
\[ \textstyle\forall~ i, u_i \leq \sum_j \uij \xij ,\]
\[\textstyle \forall~j, \sum_i \xij \leq 1, \]
\[ \xij \geq 0.\]
We now use the technology we developed in the previous section to construct the dual of this convex program. 
We let the dual variable corresponding to the constraint $ u_i \leq \sum_j \uij \xij$ be $\beta_i$ and 
the dual variable corresponding to the constraint $ \sum_i \xij \leq 1$ be $p_j$ (these will correspond to the 
equilibrium prices, hence the choice of notation). 
We also need a variable $\mu_i$ that corresponds to the variable $u_i$ in the primal program since it appears in the 
objective in the form of a  concave function, $m_i \log u_i$. We now calculate the conjugate of this function. 
Recall that if $f(x) = -\log x$ then $f^*(\mu) = -1 - \log (-\mu)$,
and if $g(x) = c f(x)$ then $g^*(\mu) = c f^*(\mu/c)$.
Therefore if $g(x) = - c \log x$ then $g^*(\mu) = -c - c\log (-\mu/c) =
c \log c - c -c\log (-\mu)$. In the dual objective, we can ignore the constant terms,
$c \log c - c$. We are now ready to write down the dual program which is as follows.
\[\textstyle \min  \sum_j p_j  - \sum_i m_i \log (-\mu_i) \text{ s.t}\]
\[\textstyle \forall~i,j, p_j \geq \uij\beta_i,\]
\[ \forall~i, \beta_i = - \mu_i  .\]
We can easily eliminate $\mu_i$ from the above to get the following program.

\begin{equation}\label{cp.dualeg} \textstyle\min \sum_j p_j  - \sum_i m_i \log (\beta_i) \text{ s.t}\end{equation}
\[ \forall~i,j, p_j \geq \uij\beta_i.\]
The variables $p_j$'s  actually correspond to equilibrium prices.
In fact, we can even eliminate the $\beta_i$'s by observing that in an optimum solution,
$\beta_i = \min_j \left\{ p_j/\uij \right \}$. This gives a convex (but not strictly convex)
function of the $p_j$'s that is minimized at equilibrium. Note that this is an unconstrained\footnote{
Although with some analysis, one can derive that the optimum solution satisfies that $p_j \geq 0$, and 
$\sum_j p_j = \sum_i m_i$, the program itself has no constraints.} minimization. 
The function is as follows

\[\min \sum_j p_j  - \sum_i m_i \log (\min_j \left\{ p_j/\uij \right \}).\]
It would be interesting to give an intuitive explanation for why this function is minimized at equilibrium.
Another interesting property of this function is that the (sub)gradient of this function at any price vector corresponds to 
the (set of) excess supply of the market with the given price vector. This implies that a tattonement style price update,
where the price is increased if the excess supply is negative and is decreased if it is positive, is actually 
equivalent to gradient descent. 

\noindent{\bf Note:} A convex program that is very similar to (\ref{cp.dualeg}) was also discovered 
independently by Garg \cite{jgarg08}. However it is not clear how they arrived at it, or 
if they realise that this is the dual of the Eisenberg-Gale convex program.

Going back to Convex Program (\ref{cp.dualeg}), we  write an equivalent program by taking the $\log$s in each of the constraints.

\[\textstyle \min \sum_j p_j  - \sum_i m_i \log (\beta_i) \text{ s.t}\]
\[\textstyle \forall~i,j, \log p_j \geq \log \uij + \log \beta_i.\]
We now think of $q_j = \log p_j $ and $\gamma_i = -\log \beta_i $ as the variables, and get the following convex program.

\begin{equation}\label{cp.dualeglogs}\textstyle \min \sum_j e^{q_j}  + \sum_i m_i \gamma_i \text{ s.t}\end{equation}
\[ \forall~i,j, \gamma_i + q_j  \geq \log \uij.\]
We now take the dual of this program. Again, we need to calculate the conjugate of the convex function that appears in the objective, namely $e^x$. We could calculate it from scratch, or derive it from the ones we have already calculated. 
Recall that if $f(x) = e^{x-1}$, then $f^*(\mu) = \mu \log \mu$,
and if $g(x) = f(x+a)$ then $g^*(\mu) = f^*(\mu) - \mu^T a$. Thus if $g(x) = e^x = f(x+1)$ then
$g^*(\mu)= f^*(\mu) - \mu = \mu \log \mu - \mu. $ The dual variable corresponding to the constraint
$\gamma_i + q_j  \geq \log \uij$ is $\bij$ and the dual variable corresponding to $e^{q_j}$ is $p_j$ (by abuse of notation,
but it turns out that these once again correspond to equilibrium prices).
Thus we get the following convex program of \cite{shmyrev2009algorithm}, which we call $\CCP$.

\[ \textstyle\max \sum_{i,j} \bij \log \uij - \sum_j (p_j \log p_j - p_j) \text{ s.t.} \tag{\CCP}\]
\[ \textstyle\forall~j, \sum_i \bij = p_j, \]
\[ \textstyle\forall~ i, \sum_j \bij = m_i,\]
\[\bij \geq 0 ~~\forall ~i,j. \]
%

\subsection{Extensions to other markets}
The Eisenberg-Gale convex program can be generalized to capture the equilibrium of many other markets, such as markets
with Leontief utilities, or network flow markets. In fact,  \cite{JVEG} identify a whole class of such markets
whose equilibrium is captured by convex programs similar to that of Eisenberg and Gale (called {\em EG markets}). We can take the dual
of all such programs to get corresponding generalizations for the convex program (\ref{cp.dualeg}).
For instance, a Leontief utility is of the form $U_i = \min_j \left\{ \xij/\phi_{ij} \right \}$ for some given values $\phi_{ij}$. 
The Eisenberg-Gale-type convex program for Fisher markets with Leontief utilities is as follows, 
along with its dual (after some simplification as before). 

\begin{minipage}{0.40\textwidth}  
	\[ \textstyle \text{Primal: }\max \sum_{i} m_i \log u_i \text{ s.t.} \]
\[ \textstyle\forall~ i,j,  u_i \leq  \xij/\phi_{ij} ,\]
\[ \textstyle\forall~j, \sum_i \xij \leq 1, \]
\[ \xij \geq 0.\]
\end{minipage}
\hfill\vline\hfill
\begin{minipage}{0.40\textwidth}  
\[\textstyle \text{Dual: }\min \sum_j p_j  - \sum_i m_i \log (\beta_i) \text{ s.t}\]
\[ \textstyle\forall~i, \sum_j \phi_{ij} p_j = \beta_i.\]
\end{minipage}

In general for an EG-type convex program, the dual has the objective function $ \sum_j p_j  - \sum_i m_i \log (\beta_i)$
where $\beta_i$ is the minimum cost buyer $i$ has to pay in order to get one unit of utility. 
For instance, for the network flow market, where the goods are edge capacities in a network and the buyers are source-sink 
pairs looking to maximize the flow routed through the network, then $\beta_i$ is the cost of the cheapest path between the 
source and the sink given the prices on the edges. 
 
However, for some markets, it is not clear how to generalize the Eisenberg-Gale convex program, 
but the dual generalizes easily.  In each of the cases, the optimality conditions can be easily seen to be 
equivalent to equilibrium conditions. We now show some examples of this.
\subsubsection*{Quasi-linear utilities}
Suppose the utility of buyer $i$  is $\sum_j (\uij - p_j)\xij$.
In particular, if all the prices are such that $p_j > \uij$, then the buyer prefers to not be allocated any good and go
back with his budget unspent. It is easy to see that the following convex program captures the equilibrium prices for such utilities. In fact, given this convex program, one could take its dual to get an EG-type convex program as well. 
\begin{minipage}{0.35\textwidth} 
\begin{equation}\label{cp.dualegQuasilinear}\textstyle \text{Primal: }\min \sum_j p_j  - \sum_i m_i \log (\beta_i) \text{ s.t}\end{equation}
\[\textstyle \forall~i,j, p_j \geq \uij\beta_i,\]
\[ \forall~i, \beta_i \leq 1. \]
\end{minipage}
\hfill\vline\hfill
\begin{minipage}{0.35\textwidth} 
\[ \textstyle \text{Dual: }\max \sum_{i} m_i \log u_i -v_i \text{ s.t.} \]
\[\textstyle \forall~ i, u_i \leq \sum_j \uij \xij + v_i ,\]
\[ \textstyle\forall~j, \sum_i \xij \leq 1, \]
\[ \xij ,v_i \geq 0 .\]
\end{minipage}

Although this is a small modification of the Eisenberg-Gale convex program, it is not clear how one would arrive at this 
directly without going through the dual. 
\subsubsection*{Transaction costs}
Suppose that we are given, for every pair, buyer $i$ and good $j$, a transaction cost $\cij$ that the buyer has to pay per unit
of the good in addition to the price of the good. Thus the total money spent by buyer $i$ is $\sum_j (p_j + \cij) \xij$.
 \citet{CDK10} show that the following convex program captures the equilibrium prices for such markets.
\begin{equation}\label{cp.dualegQuasilinear}\textstyle \min \sum_j p_j  - \sum_i m_i \log (\beta_i) \text{ s.t}\end{equation}
\[ \textstyle\forall~i,j, p_j + \cij \geq \uij\beta_i,\]
\[\textstyle \forall~i, \beta_i \leq 1. \]

\section{Convex programs for generalizations of Fisher's model}
\label{sec.convex}

In this section, we give a convex program for each market model and a complete proof that optimal solution of the convex program gives an equilibrium price of the market.

\subsection{Market $\CM_1$: Sellers have earning limits} \label{earning_bound_section}

\paragraph{Linear utilities}

This convex program is a natural extension of program $\mathcal{CP}$ presented in Section \ref{sec.4}, 
with an additional set of constraints for sellers having earning limits:
\[ \textstyle \max \sum_{i,j} b_{ij} \log u_{ij} - \sum_{j} ( q_j \log q_j - q_j ) ~ \text{s.t.} \tag{P1} \label{cp4} \]
\begin{equation}\label{con1.1}
\textstyle
 \forall j,  \sum_i b_{ij} = q_j  ,
\end{equation}
\begin{equation}\label{con1.3}
\textstyle
\forall i,\sum_j b_{ij} = m_i  ,
\end{equation}
\begin{equation}\label{con1.2}
\textstyle
\forall j,q_j \leq \SB_j ,
\end{equation}
\begin{equation}\label{con1.4}
\textstyle
\forall i,j,b_{ij} \geq 0.
\end{equation}
Here $b_{ij}$ is the amount of money buyer $i$ spends on good $j$, and $q_j$ is the total amount of spending on good $j$. Contraint \ref{con1.2} makes sure that the spending on good $j$ does not exceed the earning limit of seller $j$. 

\begin{lemma} 
Convex program \ref{cp4} captures the equilibrium prices of market model $\mathcal{M}_1$ under linear utility functions.
\end{lemma}

\begin{proof}
Let $\lambda_j, \mu_j, \eta_i$ be the dual variables for contraints \ref{con1.1}, \ref{con1.2}, \ref{con1.3} respectively. By the KKT conditions, optimal solutions must satisfy the following:
\begin{enumerate}
\item $\forall i \in B,j \in A: \quad \log u_{ij} - \lambda_j  - \eta_i \leq 0$
\item $\forall i \in B,j \in A: \quad b_{ij} >0 \Rightarrow \log u_{ij} - \lambda_j - \eta_i = 0$
\item $\forall j \in A : \quad - \log q_j + \lambda_j - \mu_j= 0$
\item $\forall j \in A : \quad  \mu_j \geq 0$
\item $\forall j \in A : \quad  \mu_j > 0 \Rightarrow q_j =  \SB_j $
\end{enumerate}

From the first 3 conditions, we have $\forall i \in B,j \in A$ : $ \frac{u_{ij}}{q_j e^{\mu_j}} \leq e^{\eta_i}$
and if $b_{ij} > 0$ then $\frac{u_{ij}}{q_j e^{\mu_j}} = e^{\eta_i}.$
Let $p_j = q_j e^{\mu_j}$. We will show that $p$ is an equilibrium price with spending $b$. From the above observation, it is easy to see that each buyer $i$ only spends money on his maximum bang-per-buck (MBB) goods at price $p$, i.e., goods that give her maximum utility per unit money spent. We also have to check that an optimal solution given by the convex program satisfies the market clearing conditions. Constraint \ref{con1.3} guarantees that each buyer $i$ must spend all his money. Therefore, we only have to show that the amount seller $j$ earns is the mininum between $p_j$ and $\SB_j$. If $q_j = \SB_j$ and $q_j \leq q_j e^{\mu_j} = p_j$. If $q_j < \SB_j$ then $\mu_j = 0$ and $p_j = q_j < \SB_j$. Thus, in both cases, $q_j = \min (p_j ,\SB_j)$ as desired.
\end{proof}


\paragraph{Spending constraint utilities}
The convex program for model $\CM_1$ under spending constraint utility functions is as follows:
\begin{equation}\label{cp5}\textstyle
\max \sum_{i,j,l} b^l_{ij} \log u^l_{ij} - \sum_{j} ( q_j \log q_j - q_j ) ~ \text{s.t.} \tag{P2}
\end{equation}
\begin{equation} \label{con5.1}\textstyle
\forall j,\sum_{i,l} b_{ij}^l = q_j,
\end{equation}
\begin{equation}\label{con5.3}\textstyle
\forall i, \sum_{j,l} b_{ij}^l = m_i,
\end{equation}
\begin{equation} \label{con5.4}\textstyle
\forall i,j, l \in S,b_{ij}^l \leq  B_{ij}^l,
\end{equation}
\begin{equation}\label{con5.2} \textstyle
\forall j,q_j \leq \SB_j,
\end{equation}
\begin{equation} \label{con5.5}\textstyle
\forall i,j, l\in S,b_{ij}^l \geq 0. 
\end{equation}
Here $b_{ij}^l$ is the amount of money buyer $i$ spends on good $j$ under segment $l$, and $q_j$ is the total amount of spending on good $j$.

\begin{lemma} 
	\label{lem:cpm1sc}
	Convex program \ref{cp5} captures the equilibrium prices of market model $\mathcal{M}_1$ under spending constraint utility function.
\end{lemma}
\begin{proof}
	Let $\lambda_j, \mu_j, \eta_i, \gamma_{ijl}$ be the dual variables for contraints \ref{con5.1}, \ref{con5.2}, \ref{con5.3}, \ref{con5.4} respectively. By the KKT conditions, optimal solutions must satisfy the following:
	\begin{enumerate}
		\item $\forall i \in B,j \in A, l \in S: \quad \log u^l_{ij} - \lambda_j  - \eta_i - \gamma_{ijl} \leq 0$
		\item $\forall i \in B,j \in A, l \in S: \quad b^l_{ij} >0 \Rightarrow  \log u^l_{ij} - \lambda_j  - \eta_i - \gamma_{ijl} = 0$
		\item $\forall j \in A : \quad - \log q_j + \lambda_j - \mu_j= 0$
		\item $\forall j \in A : \quad  \mu_j \geq 0$
		\item $\forall j \in A : \quad  \mu_j > 0 \Rightarrow q_j =  \SB_j $
		\item $\forall i \in B,j \in A, l \in S: \quad \gamma_{ijl} \geq 0$
		\item $\forall i \in B,j \in A, l \in S: \quad \gamma_{ijl} >0 \Rightarrow b_{ij}^l =  B_{ij}^l  $
		\end{enumerate}
		
		Let $p_j = q_j e^{\mu_j}$. We will prove that $p$ is an equilibrium price with spending $b$. The second KKT condition says that for a fixed pair of buyer $i$ and good $j$, $b_{ij}^l > 0$ implies 
		\[ \textstyle \frac{u_{ij}^l}{e^{\gamma_{ijl}}} = e^{\lambda_j} e^{\eta_i} \]
		Therefore, the ratio $u_{ij}^l/ e^{\gamma_{ijl}}$ is the same for every segment $l$ under which $i$ spends money on $j$. From KKT condition 7, $\gamma_{ijl} > 0$ implies $b_{ij}^l =  B_{ij}^l$. It follows that for each good $j$, $i$ must finish spending money on a segment with higher rate before starting spending money on a segment with lower rate. 
		
		From the first 3 KKT conditions, we have:
		\[\textstyle \frac{u_{ij}^l}{q_j e^{\gamma_{ijl}} e^{\mu_j}} \leq e^{\eta_i}\]
		and equality happens when $b_{ij}^l > 0$. For every segment that $i$ can still spend money on, $b_{ij}^l$ must be less than $B_{ij}^l$, and thus $\gamma_{ijl} = 0$. Therefore, for every $j$ and $l$ such that  $B_{ij}^l > b_{ij}^l > 0$, we have 
		\[\textstyle \frac{u_{ij}^l}{p_j} = \frac{u_{ij}^l}{q_j e^{\mu_j}} = e^{\eta_i}\]
		and this ratio $\frac{u_{ij}^l}{p_j}$ is maximized among all segments that $i$ can spend money on, i.e. segments such that $b_{ij}^l < B_{ij}^l$. Therefore, we can conclude that each buyer $i$ is spending according to his best spending strategy. 
		
		By complementary slackness condition, if $q_j  < \SB_j$ then $\mu_i = 0$ and $q_j = p_j$.  Otherwise, if $p_j = \SB_j$ then $q_j \leq p_j$. Therefore, in this model, the amount seller $j$ earns is the minimum between $\SB_j$ and $p_j$.
\end{proof}

\subsection{Market $\CM_2$: Buyers have utility limits} \label{utility_bound_section}

\paragraph{Linear utilities}
The convex program for the linear utility with buyers having utility limits is a natural extension of the Eisenberg-Gale program: 
\begin{equation}\label{cp1} 
\textstyle
\max \sum_{i} m_i \log u_i ~ \text{s.t.} \tag{P3} \\
\end{equation}
\begin{equation}\label{con2.1}
\textstyle
\forall i , \sum_{j} x_{ij} u_{ij} = u_i ,
\end{equation}
\begin{equation} \label{con2.2}
\textstyle
\forall i  , u_i \leq \UB_i ,
\end{equation}
\begin{equation}\label{con2.3}
\textstyle
\forall j  ,\sum_i x_{ij} \leq 1,
\end{equation}
\begin{equation}\label{con2.4}
\textstyle
\forall i,j, x_{ij} \geq 0. 
\end{equation}
In this program, $x_{ij}$ is the amount of good $j$ allocated to buyer $i$, and $u_i$ is the amount of utility that buyer $i$ obtains. Contraint \ref{con2.2} guarantees that the amount of utility buyer $i$ gets does not exceed his utility limit $\UB_i$. 

\begin{lemma}
	\label{lem:cpm2lin} 
Convex program \ref{cp1} captures the equilibrium prices of market model $\mathcal{M}_2$ under linear utility function.
\end{lemma}

\begin{proof}
	Let $\lambda_i, \mu_i, p_j$ be the dual variables for contraints \ref{con2.1}, \ref{con2.2}, \ref{con2.3} respectively. By the KKT conditions, optimal solutions must satisfy the following:
	\begin{enumerate}
		\item $\forall i \in B,j \in A: \quad -\lambda_i u_{ij}  - p_j \leq 0$
		\item $\forall i \in B,j \in A: \quad x_{ij} >0 \Rightarrow -\lambda_i u_{ij}  - p_j  = 0$
		\item $\forall i \in B : \quad \frac{m_i}{u_i} + \lambda_i - \mu_i= 0$
		\item $\forall i \in B : \quad  \mu_i \geq 0$
		\item $\forall i \in B : \quad  \mu_i > 0 \Rightarrow u_i =  \UB_i $
		\item $\forall j \in A : \quad  p_j \geq 0$
		\item $\forall j \in A : \quad  p_j > 0 \Rightarrow \sum_i x_{ij} =  1 $
	\end{enumerate}
	
	From the first 3 conditions, we have $\forall i \in B,j \in A$ : 
	$ \frac{u_{ij}}{p_j} \leq \frac{u_i}{m_i  - \mu_i u_i}$
	and if $x_{ij} > 0$ then
	$\frac{u_{ij}}{p_j} = \frac{u_i}{m_i - \mu_i u_i}.$
	
	We will show that $p$ is an equilibrium price with allocation $x$. From the above observation, it is easy to see that each buyer $i$ only spends money on his MBB goods at price $p$.
	Moreover, we know that if $p_j >0$ then good $j$ must be fully sold. Therefore, the only remaining thing to prove is that at price $p$ each buyer either spends all his money or attains his utility limit. If $u_i = \UB_i$ then buyer $i$ reaches his utility limit and the amount of money he spends is $m_i - \mu_i \UB_i$, which is at most $m_i$. If $u_i < \UB_i$ then $\mu_i = 0$ and the amount of money he spends is $m_i$.
\end{proof}

\paragraph{Leontief utilities}
The convex program for the Leontief utility model is as follows:
\begin{equation}\label{cp2}\textstyle
\max  \sum_{i} m_i \log u_i ~ \text{s.t.} \tag{P4} 
\end{equation}
\begin{equation}\label{con3.1}\textstyle
\forall i,j, u_i \phi_{ij} = x_{ij},
\end{equation}
\begin{equation}\label{con3.2} \textstyle
\forall i, u_i \leq \UB_i ,
\end{equation}
\begin{equation} \label{con3.3}\textstyle
\forall j, \sum_i x_{ij} \leq 1\\
\end{equation}
\begin{equation} \label{con3.4}\textstyle
\forall i,j,x_{ij} \geq 0.
\end{equation}
\begin{lemma} 
	\label{lem:cpm2leontief}
	Convex program \ref{cp2} captures the equilibrium prices of market model $\mathcal{M}_2$ under Leontief utility function.
\end{lemma}

\begin{proof}
	Let $\lambda_{ij}, \mu_i, p_j$ be the dual variables for contraints \ref{con3.1}, \ref{con3.2}, \ref{con3.3} respectively. By the KKT conditions, optimal solutions must satisfy the following:
	\begin{enumerate}
		\item $\forall i \in B,j \in A: \quad -\lambda_{ij}  - p_j \leq 0$
		\item $\forall i \in B,j \in A: \quad x_{ij} >0 \Rightarrow -\lambda_{ij}  - p_j = 0$
		\item $\forall i \in B : \quad \frac{m_i}{u_i} + \sum_j \lambda_{ij} \phi_{ij} - \mu_i= 0$
		\item $\forall i \in B : \quad  \mu_i \geq 0$
		\item $\forall i \in B : \quad  \mu_i > 0 \Rightarrow u_i =  \UB_i $
		\item $\forall j \in A : \quad  p_j \geq 0$
		\item $\forall j \in A : \quad  p_j > 0 \Rightarrow \sum_i x_{ij} =  1 $
	\end{enumerate}
	
	Notice that in this model, we may assume that $u_i > 0$ for all $i \in B$. It follows from constraint $\ref{con3.1}$ that $x_{ij} = 0$ if and only if $\phi_{ij} = 0$. From the second KKT condition, we know that if $\phi_{ij} >0$, we must have $ \lambda_{ij} = -p_j$. Substituting in the third condition we have:
	\[\textstyle  \frac{m_i}{u_i}  - \mu_i = \sum_j p_j \phi_{ij} \]
	Therefore, 
	\[\textstyle m_i - \mu_i u_i = \sum_j p_j \phi_{ij} \frac{x_{ij}}{\phi_{ij}} = \sum_j p_j x_{ij}\]
	It follows that $m_i - \mu_i u_i$ is actually the amount of money that buyer $i$ spends. By complementary slackness condition, if $u_i  < \UB_i$ then $\mu_i = 0$ and $i$ spends all his budget. Otherwise, if $u_i = \UB_i$ then $m_i - \mu_i u_i \leq m_i$. Therefore, in this model, a buyer $i$ either spends all his budget or attains his utility limit. Moreover, we know that if $p_j >0$ then good $j$ is fully sold. Thus, $p$ is an equilibrium price with allocation $x$.
\end{proof}

\paragraph{CES utilities}
The convex program for the CES utility model with parameter $\rho$ is as follows: 
\begin{equation} \label{cp3}\textstyle
\max \sum_{i} m_i \log u_i ~\text{s.t.} \tag{P5}
\end{equation}
\begin{equation} \label{con4.1}\textstyle
\forall i, u_i = \left( \sum u_{ij} x_{ij}^{\rho} \right)^{\frac{1}{\rho}},
\end{equation}
\begin{equation} \label{con4.2}\textstyle
\forall i ,  u_i \leq \UB_i,
\end{equation}
\begin{equation} \label{con4.3}\textstyle
\forall j, \sum_i x_{ij} \leq 1,
\end{equation}
\begin{equation} \label{con4.4}\textstyle
\forall i,j, x_{ij} \geq 0.
\end{equation}

Notice that in this model, $\partial u_i / \partial x_{ij} = u_i^{1-\rho}u_{ij} x_{ij}^{\rho -1}$ has the same term $u_i^{1-\rho}u_{ij}$ for all $x_{ij}$'s. Moreover, $\partial u_i / \partial x_{ij}$ decreases when  $x_{ij}$ increases. It follows that the best spending strategy for a buyer $i$ is to start with $x_{ij} = 0 \quad \forall j \in A$ and spend money on goods $j$ that maximize the ratio $\frac{\partial u_i / \partial x_{ij} }{p_j}$ at every point. At the end of the procedure, all goods $j$ such that $x_{ij} >0$ will have the same value for $\frac{\partial u_i / \partial x_{ij} }{p_j}$, and that value is the maximum over all goods.

\begin{lemma} 
	\label{lem:cpm2ces}
	Convex program \ref{cp3} captures the equilibrium prices of market model $\mathcal{M}_2$ under CES utility function.
\end{lemma}

\begin{proof}
	Let $\lambda_{i}, \mu_i, p_j$ be the dual variables for contraints \ref{con4.1}, \ref{con4.2}, \ref{con4.3} respectively. By the KKT conditions, optimal solutions must satisfy the following:
	\begin{enumerate}
		\item $\forall i \in B,j \in A: \quad -\lambda_{i} u_i^{1-\rho}u_{ij} x_{ij}^{\rho -1}  - p_j \leq 0$
		\item $\forall i \in B,j \in A: \quad x_{ij} >0 \Rightarrow -\lambda_{i} u_i^{1-\rho}u_{ij} x_{ij}^{\rho -1}  - p_j = 0$
		\item $\forall i \in B : \quad \frac{m_i}{u_i} + \lambda_{i}  - \mu_i= 0$
		\item $\forall i \in B : \quad  \mu_i \geq 0$
		\item $\forall i \in B : \quad  \mu_i > 0 \Rightarrow u_i =  \UB_i $
		\item $\forall j \in A : \quad  p_j \geq 0$
		\item $\forall j \in A : \quad  p_j > 0 \Rightarrow \sum_i x_{ij} =  1 $
	\end{enumerate}
	
	We will prove that $p$ is an equilibrium price with allocation $x$.  From the first there KKT conditions, we have
	\[\textstyle \frac{u_i^{1-\rho}u_{ij} x_{ij}^{\rho -1}}{p_j} \leq \frac{u_i}{m_i - \mu_i u_i} \]
	and equality happens when $x_{ij} > 0$. Therefore, $x$ is in agreement with the best spending strategy of the buyers, which says that for each buyer $i$, if $x_{ij} >0$ then  $\frac{\partial u_i / \partial x_{ij} }{p_j}$ is maximized over all $j$'s. Moreover, we can see that $m_i - \mu_i u_i$ is the amount of money buyer $i$ spends.  By complementary slackness condition, if $u_i  < \UB_i$ then $\mu_i = 0$ and $i$ spends all his budget. Otherwise, if $u_i = \UB_i$ then $m_i - \mu_i u_i \leq m_i$. Therefore, in this model, a buyer $i$ either spends all his budget or attains his utility limit. Moreover, we know that if $p_j >0$ then good $j$ is fully sold. Thus, $p$ is an equilibrium price with allocation $x$.
\end{proof}


\section{Existence and Uniqueness}

In this section, we study the existence and the uniqueness of equilibria for the market models. 
For model $\mathcal{M}_1$, we show a necessary and sufficient condition for the existence of an equilibrium. The condition works for both cases of linear utility and spending constraint utility. On the uniqueness side, we show the spending vector, $q=(q_1,\ldots,q_m)$ where $q_j$ is the money spent on good $j$, is unique. 
For model $\mathcal{M}_2$, we show that an equilibrium always exists for all utility functions we mentioned in the previous section. On the uniqueness side, the utility vector is unique.

Despite the fact that in Fisher model we have the uniqueness of price equilibrium, it is easy to see that it is not true in our generalizations.
For model $\CM_1$, consider a market with only one buyer with utility function $u(x)=x_1$ and one seller. Let $m_1=1$ and $\SB=1$. It is easy to see every price in bigger than 1 is an equilibrium price.  
For model $\CM_2$, again consider a market with only one buyer with utility function $u(x)=x_1$ and one seller. Let $\UB_1=1$ and $m_1=2$. It is easy to see every price in interval $[1,2]$ is an equilibrium price.

\subsection{Market $\CM_1$: Sellers have earning limits} 
\begin{lemma} \label{earning_existence}
 In model $\mathcal{M}_1$ under linear and spending constraint utility functions, an equilibrium price exists if and only if $\sum_j \SB_j \geq \sum_i m_i$. 
\end{lemma}
\begin{proof} An equilibrium price exists if and only if the feasible region of the convex program is not empty. We first prove that for the case of linear utility function, the program is feasible if and only if $\sum_j \SB_j \geq \sum_i m_i$. If $\sum_j \SB_j < \sum_i m_i$ then the feasible region is empty because the set of constraints \ref{con1.1}, \ref{con1.2} and \ref{con1.3} can not be satisfied together. If $\sum_j \SB_j \geq \sum_i m_i$ then $y_{ij} = \frac{m_i \SB_j}{ \sum_j \SB_j}$ gives a feasible solution because $\sum_i y_{ij} = \SB_j \frac{\sum_i m_{i}}{\sum_j \SB_j} \leq \SB_j$ and $\sum_j y_{ij} = m_i \frac{\sum_j \SB_{j}}{\sum_j \SB_j} = m_i$. 

Similarly, we can prove that  for the case of spending constraint utility, the program is feasible if and only if $\sum_j \SB_j \geq \sum_i m_i$. If $\sum_j \SB_j < \sum_i m_i$ then the feasible region is empty because the set of constraints \ref{con5.1}, \ref{con5.2} and \ref{con5.3} can not be satisfied together. Using a similar argument as in the previous part, we can show that if the amount of money that $i$ spends on $j$ is $m_i \SB_j /  \sum_j \SB_j$ then constraints \ref{con5.1}, \ref{con5.2} and \ref{con5.3} are all satisfied. We only need to guarantee that contraint \ref{con5.4} is satisfied as well. This can be done by choosing appropriate $y_{ij}^l$'s such that 
$ \sum_l y_{ij}^l =  \frac{ m_i \SB_j}{\sum_j \SB_j} \qquad \text{and} \qquad y_{ij}^l  \leq B_{ij}^l. $
\end{proof}

\begin{lemma}
In model $\mathcal{M}_1$ under linear and spending constraint utility functions, the spending vector $q$ is unique.
\end{lemma}

\begin{proof}
Consider two distinct price equilibriums $p$ and $p'$, their corresponding spending vectors $q$ and $q'$ and their corresponding demand vectors $x$ and $x'$.
Note that $p_j \geq p'_j \Rightarrow q_j\geq q'_j$ because $q_j=x_j p_j= \min (1,\frac{\SB_j}{p_j})p_j\geq \min (1,\frac{\SB_j}{p'_j})p'_j=q'_j$. 
Consider price vector $r=(r_1,\ldots,r_m)$ where $\forall k,$ $r_k= \max (p_k,p'_k)$, its corresponding spending vector $q^r$ and its corresponding demand vectors $x^r$.
Note that by changing prices from $p$ to $r$ we may only increasing the prices. Therefore, it is easy to see under linear or spending constraint utility functions the demand of good $j$ going from prices $p$ to $r$ would not decrease if $p'_j<p_j=r_j$. 
%
Therefore, 
we have $ q^r_j=x^r_j r_j=x^r_j p_j\geq x_j p_j=q_j\geq q'_j$. We can do the same for all $j$ and show $\forall j,$ $q^r_j=max(q_j,q'_j)$.
For the sake of a contradiction suppose $\exists j$, $q_j>q'_j$ then using the later it is easy to show $\sum_j q^r_j > \sum_j q_j = \sum_j q'_j=\sum_i m_i$ which is contradiction because the money spent on goods cannot be more than the total budget. Therefore, $\forall j$ , $q_j=q'_j$ and the lemma follows.
\end{proof}

\subsection{Market $\CM_2$: Buyers have utility limits}

\begin{lemma} \label{utility_existence}
In model $\CM_2$ under linear, Leontief and CES utility functions, an equilibrium price always exists. 
\end{lemma}
\begin{proof} An equilibrium price exists if and only if the feasible region of the convex program is not empty. In \ref{cp1}, \ref{cp2} and \ref{cp3}, $x_{ij} = 0$ for all $i \in B,j \in A$ is a feasible solution. Therefore, the feasible region is not empty and an equilibrium exists.
\end{proof}
\begin{lemma}
In model $\CM_2$ under linear, Leontief and CES utility functions, the utilities of an equilibrium are unique.
\end{lemma}
\begin{proof}
In section \ref{utility_bound_section}, we showed every equilibrium correspond to an solution of a convex program with an objective function of the form $\sum_{i} m_i \log u_i$.
It is easy to see that the objective function is strictly concave.
Therefore, there is a unique vector $u$ that maximizes the objective function and the lemma follows.
\end{proof}
\section{Rationality}
\label{sec.rationality}

In this section, we prove rationality results for the case of model $\mathcal{M}_2$ under linear utility, and model $\mathcal{M}_1$ under linear utility and spending constraint utility. Specifically, we show that for those market models, a rational equilibrium exists if an equilibrium exists and all the parameters are rational numbers.

\subsection{Market $\CM_1$: Sellers have earning limits} 
\begin{lemma} \label{earning_rational}
In model $\mathcal{M}_1$ under linear utility functions, a rational equilibrium exists if $\sum_j \SB_j \geq \sum_i m_i$ and  all the parameters specified are rational numbers. 
\end{lemma}
\begin{proof}
Let $A_i$ be the set of goods that buyer $i$ spends money on, $\mathcal{A}$ be the family of $A_i$'s, and $L$ be the set of sellers reaching their earning limits. An equilibrium price $p$, the corresponding spending $y$ and inverse MBB value $\alpha$, if existed, must be a point inside the polyhedron $P(\mathcal{A},L)$ bounded by the following constraints:
\begin{align*} 
\textstyle 
\forall i \in B, \forall j \in A_i \qquad &  u_{ij} \alpha_i =  p_j \\
\textstyle 
\forall j \in A \qquad & u_{ij} \alpha_i \leq p_j \\
\textstyle 
\forall i \in B, \forall j \not \in A_i \qquad & y_{ij} = 0  \\
\textstyle 
\forall i \in B \qquad & \sum_j y_{ij} = m_i \\
\textstyle 
\forall j \in L \qquad & \sum_{i} y_{ij}= \SB_j \qquad p_j \geq \SB_j \\ 
\textstyle 
\forall j \not \in L \qquad & \sum_{i} y_{ij} = p_j \qquad p_j \leq \SB_j \\ 
\textstyle 
\forall i \in B, j \in A \qquad & y_{ij} \geq 0
\end{align*}
If an equilibrium price exists, then $\mathcal{A}$ and $L$ such that $P(\mathcal{A},L)$ is non-empty must also exist. Every point inside that non-empty polyhedron must also correspond to an equilibrium price. Since $u_{ij}$'s $m_i$'s and $\SB_j$'s are rational numbers, a vertex of $P(\mathcal{A},L)$ gives a rational equilibrium price. It then follows from Lemma \ref{earning_existence} that a rational equilibrium exists if and only if $\sum_j \SB_j \geq \sum_i m_i$.
\end{proof}


\begin{lemma} 
	In model $\mathcal{M}_1$ under spending constraint utility functions, a rational equilibrium exists 
	if $\sum_j \SB_j \geq \sum_i m_i$  and all the parameters specified are rational numbers. 
	\end{lemma}
	\begin{proof}
		For a buyer $i$ and good $j$, let $S_{ij}^{+}$ be the set of segments $l$ such that $y_{ij}^l = B_{ij}^l$, $S_{ij}^{0}$ be the set of segments such that $B_{ij}^l > y_{ij}^l > 0$, and $S_{ij}^{-}$ be the set of segments such that $y_{ij}^l = 0$. Also, let $\mathcal{S}$ be the family of all $S_{ij}^{+}, S_{ij}^{0}, S_{ij}^{-}$ sets, and $L$ be the set of sellers reaching their earning limits. An equilibrium price $p$, the corresponding spending $y$ and inverse MBB value $\alpha$, if existed, must be a point inside the polyhedron $P(\mathcal{S},L)$ bounded by the following constraints:
		\begin{align*} 
		\forall i \in B, \forall j \in A, \forall l \in S_{ij}^{+} \qquad &  u_{ij}^l \alpha_i \geq  p_j \qquad y_{ij}^l = B_{ij}^l \\
		\forall i \in B, \forall j \in A, \forall l \in S_{ij}^{0} \qquad &  u_{ij}^l \alpha_i =  p_j \qquad  0 \leq y_{ij}^l \leq B_{ij}^l \\
		\forall i \in B, \forall j \in A, \forall l \in S_{ij}^{-} \qquad &  u_{ij}^l \alpha_i \leq  p_j \qquad  y_{ij}^l =0  \\
		\forall i \in B \qquad & \sum_{j,l} y_{ij}^l = m_i \\
		\forall j \in L \qquad & \sum_{i,l} y_{ij}^l= \SB_j \qquad p_j \geq \SB_j \\ 
		\forall j \not \in L \qquad & \sum_{i,l} y_{ij}^l = p_j \qquad p_j \leq \SB_j 
		\end{align*}
		Suppose that all the parameters specified are rational numbers. Again, we can see that a rational equilibrium must also exist if an equilibrium exists. It then follows that a rational equilibrium exists if and only if $\sum_j \SB_j \geq \sum_i m_i$.
		\end{proof}

\subsection{Market $\CM_2$: Buyers have utility limits}

\begin{lemma} \label{utility_rational}
In model $\mathcal{M}_2$ under linear utility functions, a rational equilibrium exists if all the parameters specified are 
rational numbers.
\end{lemma}

\begin{proof}
 Let $A_i$ be the set of goods that buyer $i$ spends money on, $\mathcal{A}$ be the family of $A_i$'s, and $L$ be the set of buyers reaching their utility limits. An equilibrium price $p$, the corresponding spending $y$ and inverse MBB value $\alpha$, if existed, must be a point inside the polyhedron $P(\mathcal{A},L)$ bounded by the following constraints:
\begin{align*} 
\textstyle 
\forall i \in B, \forall j \in A_i \qquad &  u_{ij} \alpha_i =  p_j \\
\textstyle 
\forall j \in A \qquad & u_{ij} \alpha_i \leq p_j \\
\textstyle 
\forall i \in B, \forall j \not \in A_i \qquad & y_{ij} = 0  \\
\textstyle 
\forall j \in A \qquad & \sum_i y_{ij} = p_j \\
\textstyle 
\forall i \in L \qquad & \sum_{j} y_{ij} = \alpha_i \UB_i \qquad \sum_j y_{ij} \leq m_i \\ 
\textstyle 
\forall i \not \in L \qquad & \sum_{j} y_{ij} \leq \alpha_i \UB_i \qquad \sum_j y_{ij} = m_i  \\ 
\textstyle 
\forall i \in B, j \in A \qquad & y_{ij} \geq 0
\end{align*}
Suppose that all the parameters specified in this model are rational numbers. By a similar argument to Lemma \ref{earning_rational}, we can see that an equilibrium exists if and only if a rational equilibrium exists. It follow from Lemma \ref{utility_existence} that a rational equilibrium price must always exist if all the parameters specified are rational numbers.
\end{proof}


\section{Generalized Arrow-Hurwicz Theorem for Linear Utility}
\label{sec.ah}

In this section, we prove the generalized version of  Arrow-Hurwicz theorem for both models $\mathcal{M}_1$ and $\mathcal{M}_2$ for the case of linear utilities. First we give some notation needed for the proofs. We define a \emph{spending function} to be a function $F : [n]\times [m] \rightarrow \mathcal{R}$ such that $F(i,j)$ denotes the amount of money that buyer $i$ spends on good $j$, and a \emph{spending vector} $v$ to be an $m$-dimensional vector indicating how much money is spent on each good, that is, $v_j = \sum_i F(i,j)$.

\subsection{Market $\CM_1$: Sellers have earning limits}

\begin{theorem} Let $p$ be a equilibrium price in market model $\mathcal{M}_1$ under linear utility functions.
Then for any non-equilibrium price $q$, we have $ f_{q}^\top p > \Delta_q$ 
where $f_q$ is the excess demand vector at $q$, and $\Delta_q = f_q^\top q$ is the total excess demand value at $q$. 
\end{theorem}

\begin{proof} 
Again, we assume there are two phases during which the price/spending vectors pair change from $(p,z)$ to $(q,v)$.
\begin{enumerate}
\item In the first phase, the price vector change from $p$ to $q$. However, each agent does not spend his money on his optimal bundle at $q$. Instead, he still spends money the same way he does at equilibrium $p$. Therefore, the spending vector $z$ remains unchanged in this phase.
\item In the second phase, the price vector remains unchanged, and the spending vector changes from $z$ to $v$. Let $y = v - z$ be a vector reflecting the change between two spending vectors.  
\end{enumerate}

Our goal is to prove that $f_{q}^T p > \Delta_q$, and this is equivalent to proving 
\[\textstyle  \sum_{j=1}^n \Big( \frac{v_j}{q_j} - \text{sup}(j) \Big)p_j =  \sum_{j=1}^n \Big( \frac{z_j+ y_j}{q_j} - \text{sup}(j)  \Big)p_j = \sum_{j=1}^n \Big( \frac{z_j}{q_j} - \text{sup}(j)  \Big)p_j + \sum_{j=1}^n \frac{y_j p_j}{q_j} >  \Delta_q\]
Here sup$(j)$ is the amount of good that seller $j$ is willing to sell under price $q$.

First we prove that $ \sum_{j=1}^n \Big( \frac{z_j}{q_j} - \text{sup}(j)  \Big)p_j > \Delta_q$. We may assume that the goods are arranged so that  
\begin{itemize}
\item  $1 \leq  j \leq l: \quad  p_j \geq \SB_j \quad q_j \geq \SB_j$ 
\item  $l+1 \leq  j \leq k: \quad  p_j \geq \SB_j \quad q_j \leq \SB_j$ 
\item  $k+1 \leq  j \leq k+s:\quad  p_j \leq \SB_j \quad q_j \geq \SB_j$ 
\item  $k+s+1 \leq  j \leq n: \quad p_j \leq \SB_j \quad q_j \leq \SB_j$ 
\end{itemize}

Since $p$ is an equilibrium price, the total budget of all buyers is 
$\sum_{j=1}^k \SB_j + \sum_{j=k+1}^n p_j. $
The total value of goods that sellers willing to sell at price $q$ is:
\[\textstyle  \sum_{j=1}^l \SB_j  +  \sum_{j=l+1}^k q_j + \sum_{j=k+1}^{k+s}  \SB_j  + \sum_{j=k+s+1}^n q_j\]
Therefore, the total excess demand is: 
\[\textstyle  \Delta_q =  \sum_{j=l+1}^k (\SB_j - q_j) + \sum_{j=k+1}^{k+s} (p_j - \SB_j) + \sum_{j=k+s+1}^n (p_j - q_j)  \]

Let $F(j) =  \Big( \frac{v_j}{q_j} - \text{sup}(j) \Big)p_j$, we compute $F (j)$ in 4 different cases:

\begin{enumerate}
\item{ $1 \leq j \leq l: ~ F(j) = p_j \Big( \frac{\SB_j}{q_j} - \frac{\SB_j}{q_j} \Big) = 0 $}
\item $l+1 \leq j \leq k:~ F(j) = p_j \Big(\frac{\SB_j}{q_j} - 1 \Big) = \frac{p_j(\SB_j - q_j)}{q_j} \geq \SB_j- q_j$\\
The inequality follows from the facts that $q_j \leq \SB_j$ and $q_j \leq \SB_j \leq p_j$.
\item $k+1 \leq j \leq k+s:~ F(j) = p_j \Big(\frac{p_j}{q_j} - \frac{\SB_j}{q_j} \Big) = \frac{p_j^2/\SB_j - p_j}{q_j/\SB_j} \geq p_j^2/\SB_j - p_j.$\\
The inequality follows because $p_j^2/\SB_j - p_j \leq 0$ and $q_j/\SB_j \geq 1$. 
\item $k+s+1 \leq j \leq n:~ F(j) = p_j \Big(\frac{p_j}{q_j} -1 \Big) = \frac{p_j^2 - p_j q_j}{q_j} \geq \frac{p_j^2 - p_j q_j}{p_j}  = p_j - q_j.$\\
The reason for the inequality is as follows. If $p_j \geq q_j$ then $\Big(\frac{p_j}{q_j} -1 \Big) \geq 0$, and $\frac{p_j^2 - p_j q_j}{q_j} \geq 0$, therefore increasing the denominator decreases the value of the number. If $p_j < q_j$ then $\Big(\frac{p_j}{q_j} -1 \Big) < 0$, and $\frac{p_j^2 - p_j q_j}{q_j} < 0$, therefore decreasing the denominator decreases the value of the number.
\end{enumerate}
Therefore, 
\begin{align*}
\textstyle
\sum_j F(j) &\textstyle \geq \sum^{k}_{j=l+1} (\SB_j-q_j) +  \sum^{k+s}_{j=k+1} (p_j^2/\SB_j - p_j) +  \sum^{n}_{j=k +s+1} (p_j - q_j) \\
 	   &\textstyle = \Delta_q - \sum_{j=k+1}^{k+s} (p_j - \SB_j) +  \sum^{k+s}_{j=k+1} (p_j^2/\SB_j - p_j) 
	   = \Delta_q + \sum_{j=k+1}^{k+s} (\SB_j - p_j) +   (p_j^2/\SB_j - p_j) \\
	   &\textstyle = \Delta_q  +  \sum^{k+s}_{j=k+1} \frac{ 1 }{\SB_j} (p_j^2 - 2p_j\SB_j + \SB_j^2)  
	  = \Delta_q  + \sum^{k+s}_{j=k+1} \frac{ (p_j -\SB_j)^2 }{\SB_j} 
             \geq \Delta_q 
\end{align*}
It is easy to see that if equality happens, $q$ must be an equilibrium price. Therefore, for every non-equilibrium price $q$, we must have $ \sum_{j=1}^n \Big( \frac{z_j}{q_j} - \text{sup}(j)  \Big)p_j > \Delta_q$.

Next we prove that $\sum_{j=1}^n \frac{y_j p_j}{q_j} \geq 0$. Recall that $y$ is the difference between two spendings vectors $v$ and $z$. We can break $y$ into primitive spending changes from $z$ to $v$. 
\[\textstyle  \sum_{j=1}^n \frac{y_j p_j}{q_j} = \sum_i \sum_{\text {changes } \delta_{jk} \text{ of } i} \delta_{jk} \left( \frac{p_k}{q_k} - \frac{p_j}{q_j} \right) \]
where $\delta_{jk}$ denotes a spending change from $j$ to $k$ of positive value. Since $i$ prefers $j$ to $k$ at price $p$ $\frac{u_{ij}}{p_j} \geq \frac{u_{ik}}{p_k},$ and since $i$ prefers $k$ to $j$ at price $q$ $\frac{u_{ij}}{q_j} \leq \frac{u_{ik}}{q_k}.$
It follows that for all primitive spending changes from $j$ to $k$, $\frac{p_k}{q_k} \geq \frac{p_j}{q_j}$. Therefore, 
\[\textstyle  \sum_{j=1}^n \frac{y_j p_j}{q_j} = \sum_i \sum_{\text {changes } \delta_{jk} \text{ of } i} \delta_{jk} \left( \frac{p_k}{q_k} - \frac{p_j}{q_j} \right) \geq 0\]
as desired.
\end{proof}

\subsection{Market $\CM_2$: Buyers have utility limits}
We first define a \emph{maximal spending} to be a spending function such that every buyer either reaches his utility limit or his budget limit. Note that in model $\mathcal{M}_2$, every spending function should be a maximal spending. We also define an \emph{optimal spending} to be a maximal spending in which every buyer spends money on his optimal bundle.  

First, we state a lemma needed for proving of the theorem. Roughly speaking, the lemma says that among all maximal spendings, an optimal spending results in least money spent and most utility achieved for buyers.

\begin{lemma} \label{max_opt}
Let $F$ be a maximal spending and $G$ be an optimal spending of the same price vector. For every buyer $i$:
\begin{enumerate}
\item The money $i$ spends with respect to F is at least the money $i$ spends with respect to $G$.
\item The utility $i$ gets with respect to F is at most the utility $i$ gets with respect to $G$.
\end{enumerate}
\end{lemma}

\begin{proof} 
\begin{enumerate}

\item Assume the money $i$ spends with respect to F is less than the money $i$ spends with respect to $G$ for the sake of contradiction. Since $i$ spends money on his optimal bundle with respect to $G$, the amount of utility $i$ gets with respect to $F$ must also be less than the amount of utility $i$ gets with respect to $G$. Therefore, with respect to $F$, $i$ reaches neither budget limit nor utility limit. This is a contradiction since $F$ is a maximal spending. 

\item Assume the utility $i$ gets with respect to $F$ is more than the utility $i$ gets with respect to $G$. Since $i$ spends money on his optimal bundle with respect to $G$, the amount of money $i$ spends with respect to $F$ must also be more than the amount of money $i$ spends with respect to $G$. Therefore, with respect to $G$, $i$ reaches neither budget limit nor utility limit. This is a contradiction since $G$ is a maximal spending. 
\end{enumerate}
\end{proof}

Now we can prove the main theorem

\begin{theorem} Let $p$ be a equilibrium price in market model $\mathcal{M}_2$ under linear utility functions. 
Then for any non-equilibrium price $q$, $ f_{q}^T p > \Delta_q$
where $f_q$ is the excess demand vector at $q$, and $\Delta_q = f_q^T q$ is the total excess demand value at $q$. 
\end{theorem}
\begin{proof}

We assume there are two phases during which the price/spending vectors pair change from $(p,p)$ to $(q,v)$.
\begin{enumerate}
\item In the first phase, the price vector change from $p$ to $q$. However, each agent does not spend his money on his optimal bundle at $q$. Instead, he still only spends money on the set of good he wants at equilibirum $p$. Specifically, if at price $p$ a buyer $i$ spends $x_1, \ldots , x_k$ on $k$ different goods, we break $i$ into $k$ buyers $i_1 \ldots i_k$ such that $i_t$ has budget limit $m_i x_t / \sum_{1}^k x_l$, utility limit $\UB_i x_t / \sum_{1}^k x_l$, has the same utility function as $i$ but only spends money on good $t$. The spending of $i$ in this phase is the combination of all spendings of $i_t$s at $q$. Let $F$ be the spending function of the original buyers, and $\overline{F}$ be the spending function of the divided buyers $\overline{B}$. Note that $\overline{F}$ is a maximal spending but might not be an optimal spending. Let $z$ be the spending vector with respect to $F$ (and thus also to $\overline{F}$).
\item In the second phase, the price vector remains unchanged, and the spending vector changes from $z$ to $v$. Let $G$ be the spending function of the original buyers in this phase. We know that $G$ is an optimal spending. Let $y = v - z$ be a vector reflecting the change between two spending vectors.  
\end{enumerate}

Our goal is to prove that $f_{q}^T p > \Delta_q$, and this is equivalent to proving 
\[\textstyle  \sum_{j=1}^n \Big( \frac{v_j}{q_j} - 1\Big)p_j =  \sum_{j=1}^n \Big( \frac{z_j+ y_j}{q_j} - 1\Big)p_j = \sum_{j=1}^n \Big( \frac{z_j}{q_j} - 1\Big)p_j + \sum_{j=1}^n \frac{y_j p_j}{q_j} > \Delta_q\]

First, we prove that $\sum_{j=1}^n \Big( \frac{z_j}{q_j} - 1\Big)p_j > \Delta_q$. Recall that $z_j$ is the amount of money spent on good $j$ in the first phase, where a buyer is only interested in the set of goods he buys at equilibrium $p$. We break the analysis into 2 cases: 
\begin{itemize} 
\item $q_j \geq p_j$: Consider a divided buyer $i$ spending money on $j$ at equilibrium price $p$. If $i$ spends all his budget at equilibrium, his spending on $j$ will remain unchanged in this phase. If $i$ reaches his utility limit at equilibrium, his spending on $j$ will inrease by a factor of at most $q_j / p_j$. It follows that $z_j \leq q_j$. 
\item $q_j < p_j$:  Consider a divided buyer $i$ spending money on $j$ at equilibrium price $p$. If $i$ is at his budget limit at equilibrium, his spending on $j$ may decrease by a factor of at least $q_j / p_j$. If $i$ reaches his utility limit at equilibrium, his spending on $j$ will decrease by a factor of exactly $q_j / p_j$. It follows that $z_j \geq q_j$. 
\end{itemize} 
Therefore, in both cases $\frac{z_j - q_j}{q_j/p_j} \geq z_j - q_j$. We have
\[\textstyle  \sum_{j=1}^n \Big( \frac{z_j}{q_j} - 1\Big)p_j = \sum_{j=1}^n \frac{z_j - q_j}{q_j/p_j} \geq \sum_{j=1}^n ( z_j - q_j )= \sum_{j=1}^n z_j - \sum_{j=1}^n q_j\] 
Since $G$ is an optimal spending with respect to the original buyers and the utility function is linear, $G$ can be translated into a corresponding optimal spending $\overline{G}$ with respect to the divided buyers $\overline{B}$.  Note that $ \sum_{j=1}^n v_j$ is a total optimal spending with respect to $G$ and therefore also with respect to $\overline{G}$, and $\sum_{j=1}^n z_j$ is a total maximal spending with respect to $\overline{F}$.  Since $\overline{F}$ is a feasible spending and $\overline{G}$ is an optimal spending of the same price $q$, using the first part of Lemma \ref{max_opt}, we have $\sum_{j=1}^n z_j \geq \sum_{j=1}^n v_j$. Therefore, 
$\textstyle  \sum_{j=1}^n \Big( \frac{z_j}{q_j} - 1\Big)p_j \geq \sum_{j=1}^n v_j - \sum_{j=1}^n q_j = \Delta_q$.
Moreover, it can be seen that if equality happens, $q$ must also be an equilibrium price. Therefore,  for every non-equilibrium price $q$, we must have $\sum_{j=1}^n \Big( \frac{z_j}{q_j} - 1\Big)p_j > \Delta_q$.

Now we prove that $\sum_{j=1}^n \frac{y_j p_j}{q_j} \geq  0$ by analyzing over primitive spending changes. Since $\overline{F}$ is a maximal spending and $\overline{G}$ is an optimal spending of the same price, from the second part of Lemma \ref{max_opt}, we know that the amount of utility each divided buyer gets in $\overline{G}$ is at least as much as the amount he gets in $\overline{F}$. Consider a divided buyer $i \in \overline{B}$. In $\overline{F}$, $i$ spends money on a single good $j$. In $\overline{G}$, assume that $i$ spends money on $k_1, \ldots , k_l$. We can break the spending change of $i$ from phase 1 to phase 2 into primitive changes $j \rightarrow k_1, \ldots, j \rightarrow k_t, \ldots , j \rightarrow k_l$. The notation $j \rightarrow k$ means instead of spending $\delta_{j}$ on $j$ according to $\overline{F}$, $i$ spends $\delta_{k}$ on $k$ according to $\overline{G}$ and attains at least as much utility. Let $u_{ij}$ and $u_{ik}$ be the amount of utility $i$ gets from 1 unit of good $j$ and $k$ respectively. Since $i$ prefers $j$ to $k$ at equilibrium price $p$, we have  $ \frac{u_{ij}}{p_j}\geq \frac{u_{ik}}{p_k} $ since the amount of utility he gets by spending $\delta_j$ on $j$ at price $q_j$ is at most the amount of utility he gets by spending $\delta_k$ on $k$ at price $q_k$,
$ u_{ij} \frac{\delta_j}{q_j} \leq u_{ik} \frac{\delta_k}{q_k}. $
Therefore, $\frac{p_k}{p_j} \geq \frac{u_{ik}}{u_{ij}} \geq \frac{\delta_j q_k }{ \delta_k q_j}.$
It follows that $\delta_k \frac{p_k}{q_k} - \delta_j \frac{p_j}{q_j} \geq 0.$
Therefore, summing over all divided buyers
\[\textstyle \sum_{j=1}^n \frac{y_j p_j}{q_j} = \sum_i \sum_{\text {changes }j \rightarrow k \text{ of } i} \delta_k \frac{p_k}{q_k} - \delta_j \frac{p_j}{q_j} \geq 0\]\end{proof}

%
%
%
%
%

%
%
%

\section{Polynomial Time Computation of Rational Equilibrium Prices via Ellipsoid Method}
\label{sec:ellipsoid}
\subsection{Overview}

First, we give an overview of the ellipsoid method. Suppose we want to find a point in a bounded polyheron $P$ and have access to a seperation oracle that can answer the question of whether a point $z$ is in $P$ or not, and give a seperating hyperplane in the latter case. The ellipsoid method works as follows. We first start with an initial ellipsoid that is guarantee to contain the entire polyhedron $P$.  We then call the seperation oracle on the center $z$ of the ellipsoid. If $z$ is in $P$ then we found a point in $P$ as desired. If $z$ is not in $P$ then the oracle returns a seperating hyperplane such that $P$ and $z$ are on the opposite sides of that hyperplane. Note that this seperating hyperplane cuts our ellipsoid into two half-ellipsoids, one of them contains $P$ and the other contains $z$. We then find another ellipsoid enclosing the half-ellipsoid that contains $P$ and recurse on that ellipsoid. The algorithm stops when we find a point inside $P$ or when the volume of the bounding ellipsoid becomes small enough and we are able to claim that there is no point in $P$. In this section, we apply the ellipsoid method to find an equilibrium of our market models under linear utility function. Specifically, we show how to check if a price is an equilibrium or not in polynomial time, implement a polynomial-time seperation oracle and form an initial ellipsoid. 

For the running time analysis, we restate the Theorem 12 from \cite{Jain}, which we utilize to demonstrate that a separation oracle can be used compute equilibrium solutions in polynomial time if a rational solution exists.
\begin{theorem} Given a convex set via a strong separation oracle with a guarantee that the set contains a point with binary encoding length at most $\phi$, a point in the convex set can be found in polynomial time.
\end{theorem}
This theorem is proved using ellipsoid method and simultaneous diophantine approximation, and we refer the readers to the paper \cite{Jain} for a detailed proof.

\subsection{Checking if a given price is an equilibrium price} 
Given a price $p$, the \emph{MBB graph} is a directed bipartite graph with directed edge $(i,j)$ between buyer $i$ and good $j$ if and only if $j$ is an MBB good of $i$ at price $p$. We can build a directed network as follows: assign capacity of infinity to all edges in the MBB graph; introduce a source vertex $s$ and a directed edge from $s$ to every $i \in B$ with capacity equal to the amount that $i$ is willing to spend; introduce a sink vertex $t$ and a directed edge from every $j \in A$ to $t$ with capacity equal to the value of good that $j$ is willing to sell. After that, checking if a given price is an equilibrium price can be done via one $s-t$ max-flow computation in the network. 

\subsection{Seperation oracle} 

The generalized version of Arrow-Hurwicz theorem in Section \ref{sec.ah} gives us a simple way to implement a seperation oracle in polynomial time. The theorem says that for any non-equilibrium price $q$, the half-plane $f_q^T x > \Delta_q$, where $f_q$ is the excess demand function and $\Delta_q =f_q^T q$ is the total excess spending with respect to $q$, contains all equilibrium prices. Therefore, the hyperplane $f_q^T x = \Delta_q$ can serve as a seperating hyperplane, and since we can compute $f_q$ and $\Delta_q$ in polynomial time, we have a polynomial-time seperation oracle. 

\subsection{Bounding box} 

We choose the initial ellipsoid to be the enclosing ball of an $n$ dimensional hypercube. The hypercube is guaranteed to contain least one equilibrium price. 

\subsubsection{Market $\CM_1$: Sellers have earning limits} 
Consider an equilibrium price $p$ in this market. If $p_j > \SB_j$ for all good $j$ in the market, we can scale prices of all goods down by the same factor so that there is one good $k$ with $p_k = \SB_k$. Therefore, we may assume that there exists some $k$ such that $p_k \leq \SB_k$. For a good $j$, let $i$ be a buyer buying $j$ at price $p$. We have 
\[ \frac{u_{ij}}{p_j}  \geq  \frac{u_{ik}}{p_k} \geq \frac{u_{ik}} {\SB_k} \]
and 
\[ p_j \leq \frac{u_{ij} \SB_k}{u_{ik}} \]
It follows that the hypercube $\{p:0 \leq  p_j \leq \max_{i,k} {\frac{u_{ij}\SB_k}{u_{ik}}} \}$ contains at least one equilibrium price.

\subsubsection{Market $\CM_2$: Buyers have utility limits}

In this market model, let $M$ be the total budget of all buyers. We claim that the hypercube $\{p:0 \leq  p_i \leq M \}$ contains all equilibrium prices. This is because any price outside of that cube has a coordinate $p_i > M$, and thus can not be an equilibrium price.

\section{Discussion}

An obvious and important open question is to obtain a convex program for the common generalization
of markets $\CM_1$ and $\CM_2$ in which buyers have utility bounds and sellers have earning bounds.
Despite much effort, we could not find such a convex program even for the case of linear utilities.

For market $\CM_1$, we obtained a convex program only for the case of linear utilities but not Leontief or
CES utilities since our procedure for obtaining the dual of a convex program works only if the constraints are
all linear. Can it be enhanced in suitable ways?

Finally, the Arrow-Hurwicz theorem works for utilitites satisfying weak gross substitutability. On the other
hand, its extension to our models $\CM_1$ and $\CM_2$ works only for the case of linear utilities. We believe
it should be possible to extend to utilities satisfying weak gross substitutability.

\bibliographystyle{plainnat}
\bibliography{kelly} 


\end{document}